\documentclass[twocolumn,showpacs,amsmath,amssymb,prl]{revtex4}
\usepackage{graphicx}% Include figure files
\usepackage{dcolumn}% Align table columns on decimal point
\usepackage{bm}% bold math
\usepackage{psfrag}

\begin{document}
%\mathindent10mm

\preprint{}

\title{
Z$_2$ index theorem for Majorana zero modes 
in a class D topological superconductor
}

\author{Takahiro Fukui and Takanori Fujiwara}
\affiliation{Department of Physics, Ibaraki University, Mito
310-8512, Japan}

\date{\today}

\begin{abstract}
We propose a Z$_2$ index theorem for a generic topological 
superconductor in class D. Introducing a particle-hole symmetry breaking 
term depending on a parameter and regarding it as a coordinate of 
an extra dimension,
we define the index of the zero modes and corresponding 
topological invariant for such an extended Hamiltonian. It is shown that 
these are related with the number of the zero modes of the 
original Hamiltonian modulo two.    
\end{abstract}

\maketitle

Topological invariants are useful tools in various fields in physics.
In particle physics, interesting phenomena such as chiral and gauge 
anomalies, instantons, vortices, Skyrmions and many other 
nonperturbative effects are related to the topological invariants
\cite{Current:85,Manton:04}. 
In condensed matter physics, they also play a crucial role 
in the classification of various kinds of phases of matter
\cite{ThoulessKNN:82,Kohmoto:85,Volovik:03,SchnyderRFL:08,Kitaev:08}.
Some kinds of defects \cite{Mermin:79} or solitons \cite{Manton:04}
are also classified by the use of topological numbers.

Recently, index theorems \cite{Callias:78,Weinberg:81} for fermions 
coupled with a Higgs field
in a monopole or vortex background \cite{JackiwRebbi:76,JackiwRossi:81}
have been attracting renewed interest in condensed matter physics. 
It is due to zero-energy Majorana bound states obeying non-Abelian statistics,
pioneered by Read and Green \cite{ReadGreen:00} in a $p$-wave superconductor 
and by Kitaev \cite{Kitaev:00,Kitaev:06} in quantum computations.
Extensive studies have recently predicted them
in various kinds of superfluids and superconductors
\cite{Ivanov:01,SternOM:04,SarmaNT:06,TewariSL:07,GurarieRad:07,FuKane:08,QiHRZ:09,SatoTF:09,
TanakaYN:09,Volovik:09,BergmanHur:09,SauLTS:10,TeoKane:10,Lee:09,Alicea:10,Herbut:10,
LinderTYSN:10,Herbut:10b,Nishida:10,YasuiIN:10,
Kitaev:06,TewariSS:10,SantosNCM:10,
FukuiFujiwara:10,Roy:10}.
In a special case in which the systems have enhanced symmetry (chiral symmetry), 
it has been shown \cite{FukuiFujiwara:10} that
the index theorem ensures the existence of such zero modes.
However, for generic topological superconductors, there are no index theorems
which relate zero modes with a topological invariant.

In this paper, we propose a Z$_2$ index theorem for
Majorana zero modes in a vortex of a generic topological superconductor in class D. 
We first introduce a particle-hole symmetry breaking term that depends 
continuously on a parameter. We next regard it as a coordinate of 
an extra dimension and extend the $d$-dimensional Hamiltonian to 
$(d+1)$-dimensional one with chiral symmetry. This enables us to 
define the index of the extended Hamiltonian and also the topological 
invariant corresponding to it. 
Since the index is equal to the number of zero modes of the original 
$d$-dimensional Hamiltonian {\it modulo two}, we thus have a Z$_2$ 
index theorem for the original system, 
implying that the Majorana zero mode in topological superconductors in class D
is also protected topologically.
We will give concrete calculations in two dimensional models, but 
the extension to an arbitrary higher dimensions is straightforward. 

We investigate 
the Hamiltonian proposed by Fu and Kane  \cite{FuKane:08} for the
surface states of a topological insulator with a proximity effect
of an $s$-wave superconductor,
\begin{alignat}1
{\cal H}=i\gamma^j\partial_j-\mu1\otimes\sigma^3+\gamma^{a+2}\phi_{a},
\label{Ham}%--------------------------------
\end{alignat}
where $\gamma^j$ and $\gamma^{a+2}$ ($j,a=1,2$) form a set of 
$\gamma$ matrices $\gamma^\mu$ in four dimensions satisfying 
$\{\gamma^\mu,\gamma^\nu\}=2\delta^{\mu\nu}$. We explicitly 
employ $\gamma^j=\sigma^j\otimes\sigma^3$ and $\gamma^{a+2}=1\otimes\sigma^{a}$, 
where, in the tensor product of the Pauli matrices, the former (latter) describes
the spin (Nambu) space.
The pairing potential is included as $\bm\phi=({\rm Re}\Delta(x),{\rm Im}\Delta(x))$.
Provided that $\phi_1\phi_2\mu\ne0$,
the Hamiltonian (\ref{Ham}) belongs to class D, 
since it has only particle-hole symmetry 
\begin{alignat}1
C{\cal H}C^{-1}=-{\cal H},
\nonumber
\end{alignat}
where charge conjugation operator $C$ is defined by $C=i\gamma^2\gamma^3K$ 
with $K$ being complex conjugation
\cite{AltlandZirnbauer:97,SchnyderRFL:08,Kitaev:08}.
This symmetry ensures that all states with nonzero energies
appear as paired states with $\pm$ energies.
Let $\varepsilon_n$ be an eigenvalue of the Hamiltonian (\ref{Ham}) 
labeled by an integer $n$. Then, it is natural to label
the paired state with the opposite energy by $-n$, and hence,
particle-hole symmetry can be characterized by the relation,
$\varepsilon_{-n}=-\varepsilon_n$. 

\begin{figure}[htb]
\includegraphics[width=0.35\linewidth]{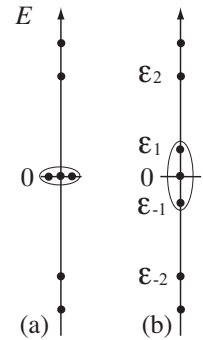}
\caption{The spectrum of the Hamiltonian in the case of $q=3$ vorticity.
(a) Three Majorana zero modes, marked by an oval, appear when $\mu=0$.
(b) Two of these get finite $\pm$ energies if a nonzero chemical potential is switched on,
but an unpaired state is protected from it, sitting exactly 
at the zero energy due to particle-hole symmetry.}
\label{f:1}%------------------------------------
\end{figure}

Let us assume that there is a vortex at the origin  
which is described by 
$\bm\phi=\Delta(r)(\cos\Theta(\theta),\sin\Theta(\theta))$, 
where $(r,\theta)$ is the polar coordinates in two dimensions 
and the phase satisfies $\Theta(2\pi)=\Theta(0)+2\pi q$ with 
$q$ being an integer.
We assume that $\Delta(0)=0$ and $\Delta(\infty)=\Delta_0>0$.
Then, it has been argued \cite{TewariSL:07,GurarieRad:07} 
that a Majorana zero mode appears for odd $q$, whereas for even $q$
no zero modes are allowed. This can be understood from the perturbation theory 
based on a simpler model with 
$\mu=0$ which belongs to class BDI due to additional chiral symmetry
\cite{AltlandZirnbauer:97,SchnyderRFL:08,Kitaev:08}. 
In this case of $\mu=0$, 
the exact $q$ zero modes wave functions can be obtained similarly 
to \cite{JackiwRossi:81}.
We can also apply the index theorem  to this special model 
\cite{Weinberg:81,FukuiFujiwara:10}, and show that the index computed by the 
exact zero modes mentioned above
and the topological invariant, i.e., the winding number of the pairing potential, indeed
coincide, both of which are given by $-q$.
This result matches the classification scheme due to Teo and Kane
\cite{TeoKane:10arXiv}. 
On the other hand, 
in the case of $\mu\ne0$, index theorems cannot apply any longer 
due to the absence of chiral symmetry. Instead, 
perturbative arguments strongly suggest that even number of zero modes 
disappear in pairs with nonzero $\pm$ energy, whereas 
an unpaired state is protected to stay at the zero energy
due to the particle-hole symmetry. 
These imply that the number of Majorana zero modes in class D 
is classified by Z$_2$, even or odd $q$ 
\cite{TewariSL:07,GurarieRad:07,FukuiFujiwara:10,TeoKane:10arXiv}.
These features are illustrated in Fig. \ref{f:1}.

\begin{figure}[htb]
\begin{tabular}{cc}
\includegraphics[width=0.4\linewidth]{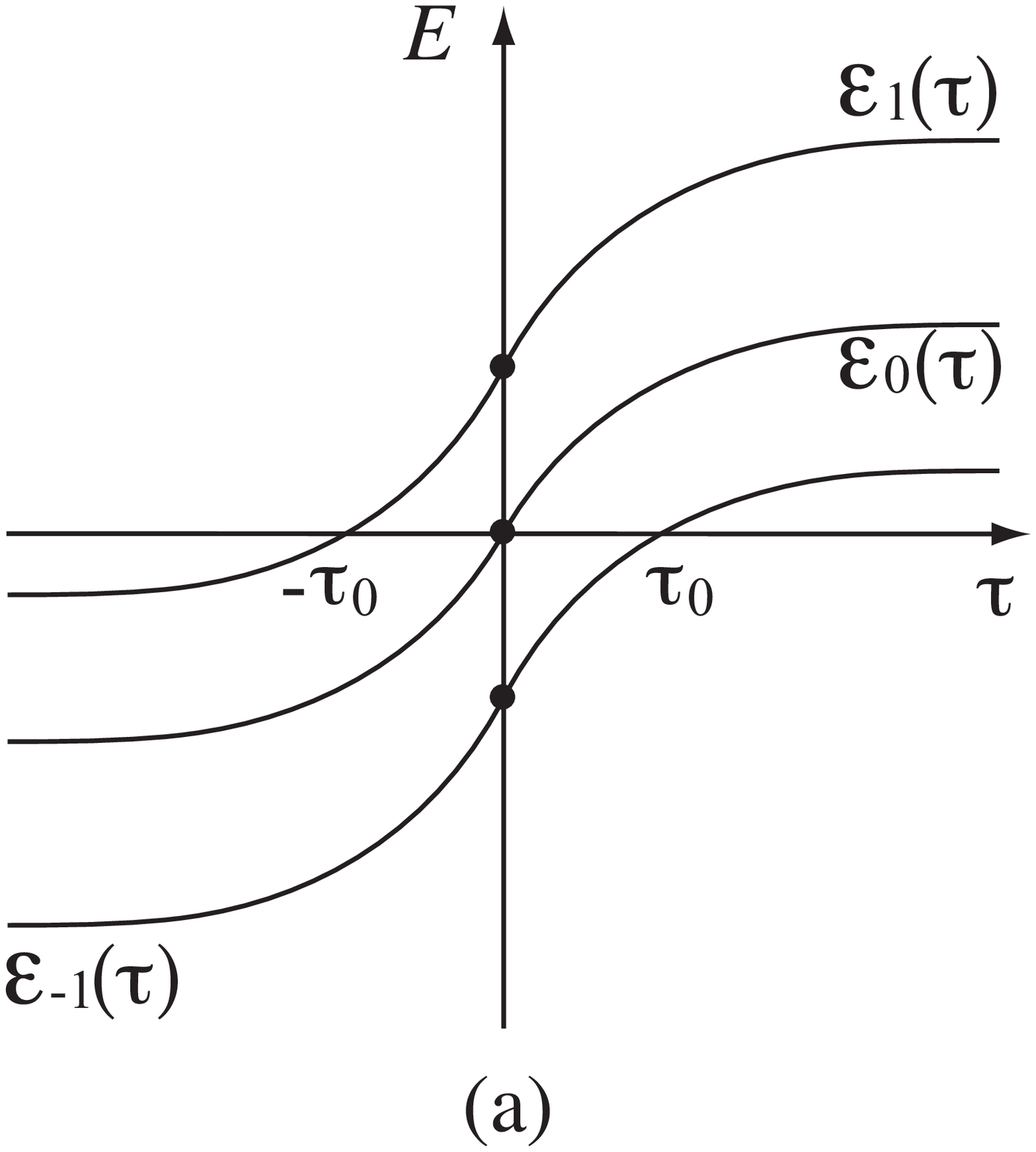}
&
\includegraphics[width=0.4\linewidth]{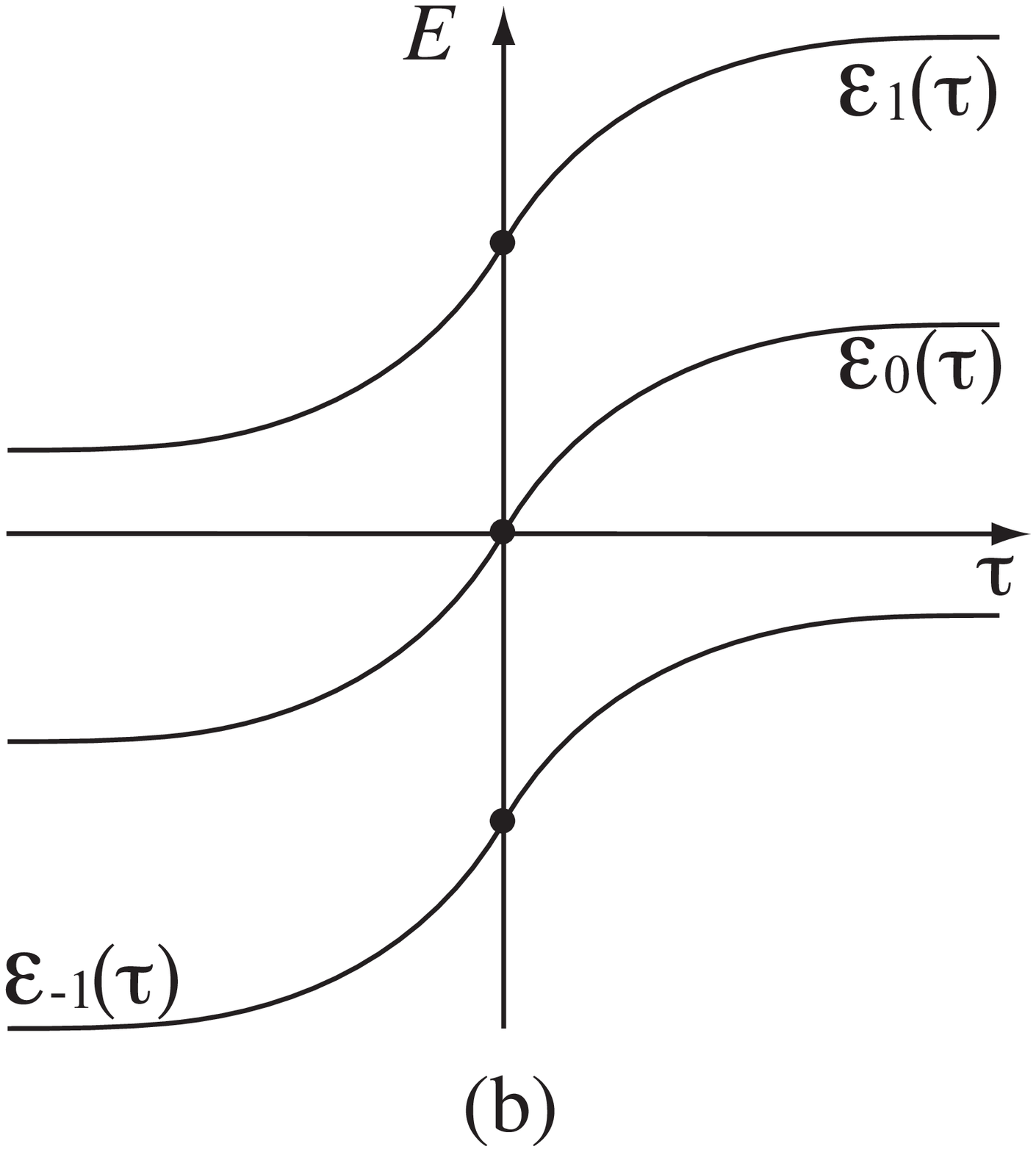}
\end{tabular}
\caption{
The spectral flow of the Hamiltonian for a generic case of $\mu\ne0$.
The $n=0$ state crosses the $E=0$ line at least $\tau=0$.
(a) If any other eigenvalue crosses it at a finite $\tau$, say, 
$\varepsilon_n(\tau_0)=0$, the spectral symmetry (\ref{PHSym2}) ensures that 
$\varepsilon_{-n}(-\tau_0)=0$.
(b) If the model parameters are changed, 
the above $\pm n$ modes may come not to cross the zero energy. Even in this case,
the index changes by {\it even} integers because of the symmetry (\ref{PHSym2}). 
}
\label{f:flow}%------------------------------------
\end{figure}

Beyond perturbative arguments, we propose an index theorem valid for
class D superconductors without chiral symmetry. To this end, 
we define, in the former part of the paper,  
an analytical index for a generic model (\ref{Ham}) with a nonzero chemical potential.
Firstly, we introduce symmetry breaking term \cite{TeoKane:10,TeoKane:10arXiv}
to the Hamiltonian (\ref{Ham}) such that 
\begin{alignat}1
{\cal H}(\tau)=i\gamma^j\partial_j-\mu1\otimes\sigma^3
+\gamma^{a+2}\phi_a-\lambda(\tau)\gamma_{\rm b},
\label{HamTau}%---------------------
\end{alignat}
where $\gamma_{\rm b}$ could be any hermitian 
matrix with $C\gamma_{\rm b}C^{-1}=\gamma_{\rm b}$, 
which may be several products of the $\gamma$ matrices. One of simpler choices is 
$\gamma_{\rm b}=\gamma_5\equiv(-i)^2\gamma^1\gamma^2\gamma^3\gamma^4$.
For simplicity and for convenience, we will restrict our discussions only to 
this case, $\gamma_{\rm b}=\gamma_5$. As for $\lambda(\tau)$, 
we assume that it is an odd function of $\tau$,
$\lambda(-\tau)=-\lambda(\tau)$, and hence $\lambda(0)=0$.
We also assume that $\lambda(\infty)=\lambda_0$ is finite.
Then, particle-hole symmetry can be generalized to
\begin{alignat}1
C{\cal H}(\tau)C^{-1}=-{\cal H}(-\tau).
\nonumber
\end{alignat}
Provided that 
the spectrum of ${\cal H}(\tau)$ is a smooth function of $\tau$,  
it can be labeled by the same quantum number $n$ for the $\tau=0$ Hamiltonian
such that
${\cal H}(\tau)\varphi_n(\tau,x)=\varepsilon_n(\tau)\varphi_n(\tau,x)$,
where $\varepsilon_n(0)=\varepsilon_n$.
From the point of view of the spectrum, the generalized particle-hole symmetry leads to 
the new relationship, 
\begin{alignat}1
\varepsilon_{-n}(-\tau)=-\varepsilon_n(\tau) .
\label{PHSym2}%-------------------------
\end{alignat}
In this sense, the states labeled by $\pm n$ can be still regarded as the paired states.
The spectral flow as a function of $\tau$ is shown in Fig. \ref{f:flow}.

Although the extended Hamiltonian (\ref{HamTau}) does not have chiral symmetry,
we can employ the spectral flow for the index theorem.
To see this, we introduce a kinetic term for the parameter $\tau$  
and define 
\begin{alignat}1
{\cal H}^{(3)}&=i\sigma^2\otimes1\partial_\tau+\sigma^1\otimes{\cal H}(\tau)
\nonumber\\
&\equiv i\Gamma^j\partial_j+\Gamma^{a+3}\phi_{a}+i\Gamma^1\Gamma^2\Gamma^6\mu ,
\label{ExtHam}%------------------------
\end{alignat}
where $j, a=1,2,3$. % and $a=1,2,3$.  
Newly defined $\Gamma$-matrices obey $\{\Gamma^\mu,\Gamma^\nu\}=2\delta^{\mu\nu}$.
${\cal H}^{(3)}$ can be regarded as a Hamiltonian defined in three dimensions 
spanned by the coordinates $x_1,x_2$ and $x_3\equiv \tau$. 
We have also introduced $\phi_3=\lambda$ and regarded it
as a component of a generalized order parameter
$\bm \phi=({\rm Re}\Delta,{\rm Im}\Delta,\lambda)$.
Note that the extended Hamiltonian (\ref{ExtHam}) has chiral symmetry 
$\Gamma_7{\cal H}^{(3)}=-{\cal H}^{(3)}\Gamma_7$, where 
$\Gamma_7=(-i)^3\Gamma^1\cdots\Gamma^6$.
Therefore, if the Hamiltonian (\ref{ExtHam}) has zero modes, they have definite chirality.

The eigenvalue equation for the zero modes is 
${\cal H}^{(3)}\Phi=0$, which is given by ${\cal H}(\tau)$ as follows:
\begin{alignat}1
\partial_\tau\Phi=\sigma^3\otimes {\cal H}(\tau)\Phi.
\label{ZerModEqu}%-------------------------
\end{alignat} 
To solve this equation, let us set $\Phi_n(\tau,x)=f_n(\tau)\varphi_n(\tau,x)$.
Then, 
$\partial_\tau\Phi_n=(\partial_\tau f_n)\varphi_n
+f_n(\partial_\tau\lambda)\partial_\lambda\varphi_n$,
since the Hamiltonian ${\cal H}(\tau)$ depends on $\tau$ only through $\lambda(\tau)$.
Provided that $\partial_\tau\lambda$ can be neglected in the adiabatic approximation  
and that $\varphi_n(\tau,x)$ is normalizable,   
it turns out that $f_n$ is given by
\begin{alignat}1
f_n^\pm(\tau)=e^{\pm \int^\tau d\tau'\varepsilon_n(\tau')}f_{\rm c}^\pm ,
\nonumber
\end{alignat}
where $f_{\rm c}$ is a constant spinor with $\Gamma_7 f_{\rm c}^\pm= \pm f_{\rm c}^\pm$.
If a given state for $n\ne0$ satisfies 
$\varepsilon_n(+\infty)>0$ and $\varepsilon_n(-\infty)<0$,
which is the case of $n=1$ and $n=-1$ modes in Fig. \ref{f:flow} (a),
the states labeled by $n$ and $-n$ are normalizable zero modes with chirality $-1$.
Likewise, if
$\varepsilon_n(+\infty)<0$ and $\varepsilon_n(-\infty)>0$,
$\pm n$ states are normalizable zero modes with chirality $+1$.
%Therefore, if we 
Now, define the index of the Hamiltonian ${\cal H}^{(3)}$ by
\begin{alignat}1
{\rm ind}\,{\cal H}^{(3)}=N_+-N_- .
\label{Ind}%-----------------------
\end{alignat}
Then, these paired zero modes contribute to the index for ${\cal H}^{(3)}$ always by two.
On the other hand, if $\varepsilon_n(\pm\infty)$ has the same sign, 
such $\pm n$ states cannot be zero modes, since the wave functions are 
not normalizable.
Therefore, these states give no contribution to the index. 
It thus turns out that $n\ne0$ modes can change the index by even integers.
Contrary to these modes, the $n=0$ mode,
which is always a zero mode of ${\cal H}^{(3)}$ unless $\varepsilon_0(\infty)=0$,
determines whether the index is even or odd.
Thus, it turns out that the number of zero modes of ${\cal H}$ in Eq. (\ref{Ham}),
which we denote as $N_0({\cal H})$, is given by the index of ${\cal H}^{(3)}$ in 
Eq. (\ref{Ind}) as 
\begin{alignat}1
N_0({\cal H})={\rm ind}\,{\cal H}^{(3)} \quad {\rm mod}~2 .
\label{NumZerMod}%-------------
\end{alignat}

So far we have discussed the index of the extended Hamiltonian (\ref{ExtHam})
and its modulo-two relationship with the number of 
zero modes of the original Hamiltonian (\ref{Ham}). 
The rest of the present paper is devoted to calculations of the corresponding
topological invariant.
For the Hamiltonian with chiral symmetry such as ${\cal H}^{(3)}$, it is possible to 
define the topological invariant equal to the index and to claim the index theorem
\cite{Callias:78,Weinberg:81,FukuiFujiwara:10}
such that
\begin{alignat}1
{\rm ind }~ {\cal H}^{(3)} &=
-\frac{1}{2}\int dS_jJ^j(x,0,\infty),
\label{IndThe1}%------------------------
\end{alignat} 
where $dS_j$ is an infinitesimal surface elements at the boundary $(x\rightarrow\infty)$
of the Euclidean space $R^3$, and $J^i$ is the axial vector current defined by
\begin{alignat}1
&J^i(x,m,M)
\nonumber\\
&=\lim_{y\rightarrow x}{\rm tr}\,\Gamma_7\Gamma^i
\left(
\frac{1}{m-i{\cal H}^{(3)}}-\frac{1}{M-i{\cal H}^{(3)}}
\right)\delta(x-y) .
\nonumber
\end{alignat}
The first term in the above parentheses becomes 
the index in the limit $m\rightarrow0$ when integrated 
over the two-dimensional surface at infinity.
The second term has been introduced as a Pauli-Villars regulator in order for the 
current to be regularized  \cite{Weinberg:81,FukuiFujiwara:10}.
It should be set $M\rightarrow\infty$ after the calculations.
When we calculate the r.h.s. of Eq. (\ref{IndThe1}), 
it may be easy to use the plane wave basis. Possible terms contributing to the index are
\begin{widetext}
\begin{alignat}1
J^i(x,0,\infty)=-\int\frac{d^3k}{(2\pi)^3}G^3
{\rm tr}\,\Gamma_7\Gamma^i
\left(-i\Gamma^jk_j-i\Gamma^{a+3}\phi_a+\Gamma^1\Gamma^2\Gamma^6\mu\right)
(K-\mu\Gamma)\Lambda(K-\mu\Gamma)\Lambda(K-\mu\Gamma) ,
\nonumber
\end{alignat}
where
\begin{alignat}1
&G^{-1}=\left[\left(\sqrt{k_1^2+k_2^2+\lambda^2}-\mu\right)^2+k_3^2+\Delta^2\right]
\left[\left(\sqrt{k_1^2+k_2^2+\lambda^2}+\mu\right)^2+k_3^2+\Delta^2\right],
\label{Pro}\\  %-------------------------
&
K=k_j^2+\phi_a^2+\mu^2,\quad
\Gamma=2i\left(\Gamma^2\Gamma^6k_1+\Gamma^6\Gamma^1k_2+\Gamma^1\Gamma^2\lambda\right),\quad 
\Lambda=i\Gamma^j\Gamma^{a+3}\partial_j\phi_a .
\nonumber
\end{alignat}
\end{widetext}

The first step of the calculations is to take the trace of the $\Gamma$-matrices by the
use of 
${\rm tr}\,\Gamma_7\Gamma^{\mu_1}\Gamma^{\nu_1}\cdots\Gamma^{\mu_3}\Gamma^{\nu_3}
=(2i)^3\epsilon^{\mu_1\nu_1\cdots\mu_3\nu_3}$.
Lengthy but straightforward calculations lead to 
\begin{alignat}1
J^i&=2^3\epsilon^{ijk}\epsilon^{abc}\phi_a\partial_j\phi_b\partial_k\phi_c
\int\frac{d^3k}{(2\pi)^3}G^2K
\nonumber\\ 
&-\delta^{i3}\mu^22^5\epsilon^{ab3}\lambda\partial_1\phi_a\partial_2\phi_b
\int\frac{d^3k}{(2\pi)^3}G^2 .\nonumber
\end{alignat}
\begin{figure}[htb]
\includegraphics[width=0.4\linewidth]{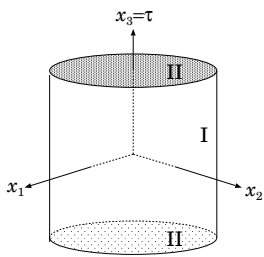}
\caption{
The cylindrical surface of the integration in Eq. (\ref{IndThe1}).
``I'' denotes the side of the cylinder $r\rightarrow\infty$, whereas
``II'' denote two discs at $\tau\rightarrow\pm\infty$. 
}
\label{f:cyl}%------------------------------------
\end{figure}
As a next step, we carry out the integration over the momentum $k_j$ in the above
and over the space coordinates $x^j$ 
in Eq. (\ref{IndThe1}). In particular in the latter integration,
since the generalized order parameter $\bm\phi$ 
depends cylindrically on the  
coordinates $(r,\theta,\tau)$, it may be natural to regard the boundary of 
$R^3$ as a cylinder illustrated in
Fig. \ref{f:cyl}. For convenience, let us divide it 
into two pieces I ($r\rightarrow\infty$) and II ($\tau\rightarrow\pm\infty$).
The calculations on these two regions are quite similar to those in \cite{Fukui:10}
when $\mu=0$. However, in the case of $\mu\ne0$, we have to choose appropriate 
$\lambda_0\equiv\lambda(\infty)$ to make the calculations valid.
To see this, 
let us set $\Delta\rightarrow\Delta_0$ and 
$\lambda\rightarrow\lambda_0$, respectively, in the regions I and II. 
Then, we see that in the region I, $G$ in Eq. (\ref{Pro})
is always finite because of finite $\Delta_0$. 
On the other hand, 
it becomes singular in the region II if $|\lambda_0|<|\mu|$. 
Indeed, in this case, $G^{-1}$ in Eq. (\ref{Pro}) vanishes at some
momentum at the core of the vortex $r=0$, where $\Delta(0)=0$.
Therefore, we restrict our discussions to the case of $|\lambda_0|\ge|\mu|$.
After some straightforward calculations, we arrive at
\begin{alignat}1
{\rm ind }~ {\cal H}^{(3)}
%\nonumber\\
&=
\frac{1}{2\mu}
\left(|\lambda_0-\mu|-|\lambda_0+\mu|\right)
\frac{1}{2\pi}\oint d\theta\epsilon^{ab}\hat\phi_a\partial_\theta\hat\phi_b
\nonumber\\
&=
-{\rm sgn}(\lambda_0)q,
\label{IndThe2}%------------------------
\end{alignat} 
where $\hat\phi^a\equiv\phi^a/\Delta_0$, the subscript $a,b$ are restricted to 
$1,2$, and the 
$\theta$-integration is over S$^1$ at $r\rightarrow\infty$ in the region II,
which becomes $q$. 
It turns out that the index is basically determined only by the vorticity $q$,
and is the same as the 
the index of ${\cal H}$ when $\mu=0$, which
has been calculated in ref. \cite{FukuiFujiwara:10}
as ${\rm ind}~{\cal H}=-q$: 
The artificially introduced
parameter $\lambda$ just change the sign of the index. 
It follows that the index is topological: It is indeed protected 
from infinitesimal changes of the three 
parameters $\mu$, $q$, and $\lambda_0$.
Eqs. (\ref{IndThe2}) and (\ref{NumZerMod}) are Z$_2$ index theorem for
a generic model in class D.

As we have mentioned, the topological invariant (\ref{IndThe2}) is invalid 
if $|\lambda_0|<|\mu|$.
In this case, the zero mode equation (\ref{ZerModEqu}) may not have 
normalizable solutions. This can be checked by the use of perturbations to
this equation.
Let us consider two possibilities of small parameters, $\mu$ and $\lambda$. 
If one regards $\mu$ as a small parameter and applies the first order 
perturbation to $\mu=0$ solutions, one has indeed one normalizable solution.
On the other hand, in the case of small $\lambda$, one can obtain the first order 
perturbative energy eigenvalue, but the first order wave function cannot be normalizable.
This implies that in the case $|\lambda|\ll|\mu|$, the relationship between
the spectral flow of ${\cal H}(\tau)$ and the zero mode of ${\cal H}^{(3)}$ does not hold.
Therefore, we conclude, from these observations, 
together with the calculations of the topological invariant, that
the present formulation of the Z$_2$ index theorem requires
$|\lambda_0|\ge|\mu|$.

\begin{acknowledgments}
The authors would like to thank H. Suzuki and M. Nitta for valuable discussions.
This work was supported in part by Grants-in-Aid for Scientific Research
(Grants No. 20340098 and No. 21540378).
\end{acknowledgments}


\begin{thebibliography}{99} %% The number "99" means that this list has
 %% more than nine items.
\bibitem{Current:85}
S. B. Treiman {\it et. al.} eds, 
{\it Current algebra and anomalies} (World Scientific Publishing Co Pte Ltd, 
Singapore, 1985).
%
\bibitem{Manton:04}
N. Manton and P. Sutcliffe,
{\it Topological Solitons} (Cambridge University Press, Cambridge, 2004).
%
\bibitem{ThoulessKNN:82}
D.J. Thouless, M. Kohmoto, M. P. Nightingale, and M. den Nijs,
Phys. Rev. Lett. {\bf 49}, 405 (1982).
%
\bibitem{Kohmoto:85}
M. Kohmoto, 
Ann. Phys. {\bf 160}, 355 (1985).
%
\bibitem{Volovik:03}
G. E. Volovik, {\it The Universe
in a Helium Droplet} (Oxford University Press, Oxford, 2003).
%
\bibitem{SchnyderRFL:08}
A. Schnyder, S. Ryu, A. Furusaki, and A. Ludwig, 
Phys. Rev. B {\bf 78}, 195125 (2008); AIP Conf. Proc. {\bf 1134}, 10 (2009).
%
\bibitem{Kitaev:08}
A. Kitaev, 
Proceedings of the L.D.Landau Memorial Conference
``Advances in Theoretical Physics'', Chernogolovka, Moscow region,
Russia, 22-26 June 2008 (unpublished).
%
\bibitem{Mermin:79}
For a review, see 
N. D. Mermin,
Rev. Mod. Phys. {\bf 51}, 591 (1979).
%
\bibitem{Callias:78}
C. Callias,
Commun. Math. Phys. {\bf 62}, 213 (1978).
%
\bibitem{Weinberg:81}
E. J. Weinberg,
Phys. Rev. D {\bf 24}, 2669 (1981).
%
\bibitem{JackiwRebbi:76}
R. Jackiw and C. Rebbi,
Phys. Rev. D {\bf 13}, 3398 (1976).
%
\bibitem{JackiwRossi:81}
R. Jackiw and P. Rossi,
Nucl. Phys. {\bf B190}, 681 (1981).
%
\bibitem{ReadGreen:00}
N. Read and D. Green,
Phys. Rev. B {\bf 61}, 10267 (2000).
%
\bibitem{Kitaev:00}
A. Kitaev,
Proceedings of the Mesoscopic and Strongly Correlated
Electron Systems Conference, Chernogolovka, Moscow
Region, Russia, 9-16 July 2000 (unpublished),
(arXiv:cond-mat/0010440).
%
\bibitem{Kitaev:06}
A. Kitaev,
Ann. Phys. {\bf 321}, 2 (2006).
%
\bibitem{Ivanov:01}
D. A. Ivanov,
Phys. Rev. Lett. {\bf 86}, 268 (2001).
%
\bibitem{SternOM:04}
A. Stern, F. von Oppen, and E. Mariani,
Phys. Rev. B {\bf 70}, 205338 (2004).
%
\bibitem{SarmaNT:06}
S. Das Sarma, C. Nayak, and S. Tewari,
Phys. Rev. B {\bf 73}, 220502 (2006).
%
\bibitem{TewariSL:07}
S. Tewari, S. Das Sarma, and D.-H. Lee,
Phys. Rev. Lett. {\bf 99}, 037001 (2007).
%
\bibitem{GurarieRad:07}
V. Gurarie and L. Radzihovsky, 
Phys. Rev. B {\bf 75}, 212509 (2007).
%
\bibitem{FuKane:08}
L. Fu and C. L. Kane,
Phys. Rev. Lett. {\bf 100}, 096407 (2008).
%
\bibitem{QiHRZ:09}
X.-L Qi, T. L. Hughes, S. Raghu, and S.-C. Zhang,
Phys. Rev. Lett. {\bf 102}, 187001 (2009).
%
\bibitem{SatoTF:09}
M. Sato, Y. Takahashi, and S. Fujimoto,
Phys. Rev. Lett. {\bf 103}, 020401 (2009).
%
\bibitem{TanakaYN:09}
Y. Tanaka, T. Yokoyama, and N. Nagaosa,
Phys. Rev. Lett. {\bf 103}, 107002 (2009).
%
\bibitem{Volovik:09}
G. E. Volovik,
Pis'ma ZhETF {\bf 90}, 639 (2009).
(arXiv:0909.3084)
%
\bibitem{BergmanHur:09}
D. L. Bergman and K. Le Hur,
Phys. Rev. B {\bf 79}, 184520 (2009).
%
\bibitem{SauLTS:10}
J. D. Sau, R. M. Lutchyn, S. Tewari, and S. Das Sarma,
Phys. Rev. Lett. {\bf 104}, 040502 (2010).
%
\bibitem{TeoKane:10}
J. C. Y. Teo and C. L. Kane,
Phys. Rev. Lett. {\bf 104}, 046401 (2010).
%
\bibitem{Lee:09}
P. A. Lee, 
arXiv:09072681.
%
\bibitem{Alicea:10}
J. Alicea, 
Phys. Rev. B {\bf 81}, 125318 (2010).
%
\bibitem{Herbut:10}
I. F. Herbut,
Phys. Rev. Lett. {\bf 104}, 066404 (2010).
%
\bibitem{LinderTYSN:10}
J. Linder, Y. Tanaka, T. Yokoyama, A. Sudbo, and N. Nagaosa,
Phys. Rev. Lett. {\bf 104}, 067001 (2010).
%
\bibitem{Herbut:10b}
I. F. Herbut,
Phys. Rev. B {\bf 81}, 205429 (2010).
%
\bibitem{TewariSS:10}
S. Tewari, J. D. Sau, and A. Das Sarma,
Annals Phys. {\bf 325}, 219 (2010).
%
\bibitem{SantosNCM:10}
L. Santos, T. Neupert, C. Chamon, and C. Mudry,
Phys. Rev. B {\bf 81}, 184502 (2010).
%
\bibitem{FukuiFujiwara:10}
T. Fukui and T. Fujiwara, 
J. Phys. Soc. Jpn. {\bf 79}, 033701 (2010).
%
\bibitem{Roy:10}
R. Roy,
arXiv:1001.2571.
%
\bibitem{Nishida:10}
Y. Nishida, 
Phys. Rev. D {\bf 81}, 074004 (2010).
%
\bibitem{YasuiIN:10}
S. Yasui, K. Itakura, and M. Nitta,
Phys. Rev. D {\bf 81}, 105003 (2010).
%
\bibitem{AltlandZirnbauer:97}
A. Altland and M. Zirnbauer,
Phys. Rev. B {\bf 55}, 1142 (1997).
%
\bibitem{TeoKane:10arXiv}
J. C. Y. Teo and C. L. Kane,
arXiv:1006.0690.
%
\bibitem{Fukui:10}
T. Fukui, 
Phys. Rev. B {\bf 81}, 214516 (2010).
%
%
\end{thebibliography}
\end{document}